\documentclass[fleqn,10pt]{wlscirep}

\usepackage[version=3]{mhchem}
\usepackage{graphicx}
\usepackage{dcolumn}
\usepackage{bm}
\usepackage{amsmath}
\usepackage{amssymb}
\usepackage{txfonts}
\usepackage{url}
\usepackage{comment}
\usepackage{tabularx}
\usepackage{endnotes}
\title{New Insight into the Ground State of FePc: A Diffusion Monte Carlo Study}
\author[1,*]{Tom Ichibha}
\author[2]{Zhufeng Hou}
\author[1,2,3]{Kenta Hongo}
\author[1]{Ryo Maezono}
\affil[1]{School of Information Science, JAIST, Nomi, Ishikawa, Japan}
\affil[2]{National Institute of Materials Science, Tsukuba, Ibaraki, Japan}
\affil[3]{PRESTO, JST, Kawaguchi, Saitama, Japan}
\affil[*]{ichibha@icloud.com}
\begin{abstract}
We have applied DMC to evaluate 
relative stability of the possible electronic 
configurations of an isolated FePc under $D_{4h}$ symmetry,
considering some fixed nodes generated from different methods.
They predict $A_{2g}$ ground state consistently, supporting 
preceding DFT studies,
with confidence overcoming the ambiguity about 
exchange-correlation (XC) functionals. 
By comparing DMC with several XC, 
we clarified the importance of the short range 
exchange to describe the relative stability.
We examined why the predicted $A_{2g}$ is excluded 
from possible ground states in the recent 
ligand field based model.
Simplified assumptions made in the 
superposition model 
are identified to give unreasonably 
less energy gain for $A_{2g}$ 
when compared with the reality. 
The state is found to have possible reasons 
for the stabilization, 
reducing the occupations from an unstable 
anti-bonding orbital, 
avoiding double occupation of a spatially localized orbital, 
and gaining exchange energy 
by putting a triplet spin pair in degenerate orbitals.
\end{abstract}
\begin{document}
\flushbottom
\maketitle
\thispagestyle{empty}

\section*{Introduction}
\label{sec:introduction}
Iron(II) Phthalocyanine (FePc)
attracts recent interests 
for its potentials in spintronics~\cite{2012KRO,2015FER, 2010TOR}
because it possesses the strong magnetic anisotropy as a 
molecular magnet.\cite{2006FIL}
It has been reported the anisotropy can be 
controlled by surrounding environments of 
molecules, such as ligands, 
polymorphs of molecular crystal structures {\it etc.}
\cite{2009TSU,2012NAK,2015FER}
These environments tune the electronic configuration 
of the central transition metal element, affecting 
its magnetic anisotropy.
Identifying the electronic configuration 
under given environments is therefore the most essential 
starting point for further understandings and applications 
of the magnetic anisotropy of these compounds,
stimulating intensive studies in this direction. 
An earlier study~\cite{1935KLE} reported 
its spin multiplicity being
in between $S=1\sim 2$.~\cite{1935KLE}
Later, Dale {\it et al.}\cite{1968DAL} 
performed magnetic susceptibility measurements 
of $\beta$-FePc, reporting that 
the system takes $S=1$ in the range, 
$T=1.25 \sim 20$~K. 
Since then, the possible configurations 
within $S=1$ have been of the interests. 
Even under this constraint, 
no consensus has been established 
about its ground state configuration. 

Experimentally, the most common targets are 
molecular crystals with lamination angles,  
$\phi=44.8^{\circ}$ ($\beta$ phase/most stable 
structure) and 
$\phi=26.5^{\circ}$ ($\alpha$ phase/quasi stable).
The most stable $\beta$ phase has been the main 
interest until 
$\alpha$-FePc was reported~\cite{1998SEL,1999YEE}
to exhibit ferromagnetic transition 
at $T_c = 5.6$~K, while $\beta$ remains paramagnetic 
until above 1~K. 
The $\alpha$ phase then attracts broader interests 
for its ferromagnetism with the spin anisotropy
lying within its molecular plane 
with unquenched orbital angular momentums.\cite{2010BAR,2015FER}
Under practical samples in experiments,
the ground state configuration of the $\alpha$ phase
has been reported as $E_g(a)$, forming a consensus.
\cite{2006FIL,2010BAR,2012KRO}

Apart from intensive discussions 
on possible factors affecting the configuration in practical
samples, such as inter-complex interactions in crystals 
\cite{2012NAK,2015BID}, the spin-orbit coupling {\it etc.}, 
\cite{1991REY,2015FER}
it is a reasonable option to start considering the simplest situation,
namely, an isolated, highly symmetric molecule.
Electronic structure calculations using DFT (density functional theory)
have hence been made for the isolated molecule.
However, even without the spin-orbit coupling,
theoretical predictions have never dropped in a consensus
as described below.
We hence target the most simplified 
question, ``{\it what is the ground state electronic 
  configuration for the ideal isolated FePc within
  the non-relativistic framework?}''. 
Most of DFT studies so far predict $A_{2g}$ 
ground state,~\cite{2001LIA,2009MAR,2012NAK} 
while a recent study\cite{2009MAR}
reports that the prediction actually depends 
on the choice of exchange-correlation (XC) 
functionals and Gaussian basis set level, getting 
both $A_{2g}$ and $B_{2g}$ as the possibility.
The same conclusion is actually confirmed in 
the present study, getting mainly $A_{2g}$, 
but sometimes $B_{2g}$ and even $E_{g}(a)$ 
depending on XC.

Another complementary approach is the 
ligand field framework.~\cite{2009MIE,2013KUZ,2015FER}
It is capable to be applied to the $D_{4h}$ isolated 
molecule, predicting $E_{g}(b)$ ground state 
for the ligand parameter choice for $\alpha$ phase.
(When the spin-orbit coupling is taken into account, 
it predicts a hybrid state between $E_{g}(b)$ and 
$B_{2g}$ as the ground state, latter of which is 
predicted as the first excited state when without 
the coupling. The magnetic anisotropy is predicted 
being perpendicular to [within] the molecular plane 
for $E_{g}(b)$ [$B_{2g}$]. 
In the hybrid state, it is within the plane 
despite the dominant state is $E_{g}(b)$.\cite{2015FER})
The original ligand field model for $D_{4h}$ 
requires three ligand parameters to identify the 
possible ground state, where $A_{2g}$ 
still remains as a possibility~\cite{2009MIE}, 
not conflicting with {\it ab initio} DFT predictions. 
In a recent work\cite{2013KUZ}, however, 
the possible ground state is specified by 
reduced two parameters and $A_{2g}$ has disappeared 
from the possibility, leading to an apparent 
contradiction to DFT predictions. 
The reduction of the freedom of parameters occurs 
when they employ the superposition model~\cite{1989NEW}
under some assumptions.

The present study targets to investigate the 
apparent discrepancy about $A_{2g}$ ground state 
between {\it ab initio} and ligand field model~\cite{2013KUZ} 
approaches.
Blocked by the ambiguity of predictions due to XC, 
this discrepancy has not well been addressed and 
investigated so far. 
To prevent the ambiguity, we applied (fixed-node)
diffusion Monte Carlo (DMC)~\cite{2001FOU}
to calibrate the XC dependence.~\cite{2012HON,2015HON,2013HON}
  Although CASSCF (complete active space self-consistent field) 
  seems a natural choice of trial nodes
  appropriate for describing the multi-reference nature in transition metals,
  a recent work suggests it is not necessarily the best for iron complexes, 
  finding some DFT trial nodes are better.~\cite{2016FUM}
  Hence we tried several trial nodes generated from
  DFT with M06, M06L, and M06-2X functionals as well as CASSCF
  (Computational details in the present study are given in
  Supplementary Information.)

We have found all the DMC predictions 
support $A_{2g}$ ground state,
being consistent with most of previous DFT calculations. 
The apparent contradiction with the ligand field 
model~\cite{2013KUZ} can be explained by further considering
the validation of the assumptions in the superposition model. 
The assumption turns out not capable to capture the 
stabilizing mechanisms of $A_{2g}$, which are clarified by 
the orbital shape/occupation analysis by the present study.

\section*{System} \label{sec:methods}
We investigate the ground state electronic configuration 
of an isolated FePc molecule under $D_{4h}$ symmetry.
While there are two preceding studies reporting the possible 
geometry, the one from X-ray diffraction of $\beta$ phase~\cite{1976KIR}
and the other from DFT geometry optimization applied to 
an isolated complex,~\cite{2008SUM}
we used the latter for the present calculation
(See Supplementary Information).
The system accommodates six electrons in 
$3d$-shells from Fe ion. 
Within the constraint of spin triplet, $S=1$, 
there are four possible configurations labeled as,
\begin{equation}
  \begin{split}
    A_{2g}&:(a_{g})^{\uparrow\downarrow}(e_{g})^{\uparrow\uparrow}(b_{2g})^{\uparrow\downarrow}
    = (d_{z^2})^{\uparrow\downarrow}(d_{xz,yz})^{\uparrow\uparrow}(d_{xy})^{\uparrow\downarrow},\\
    B_{2g}&:(a_{g})^{\uparrow}(e_{g})^{\uparrow\downarrow\uparrow\downarrow}(b_{2g})^{\uparrow}
    = (d_{z^2})^{\uparrow}(d_{xz,yz})^{\uparrow\downarrow\uparrow\downarrow}(d_{xy})^{\uparrow},\\
    E_{g}(a)&:(a_{g})^{\uparrow}(e_{g})^{\uparrow\downarrow\uparrow}(b_{2g})^{\uparrow\downarrow}
    = (d_{z^2})^{\uparrow}(d_{xz,yz})^{\uparrow\downarrow\uparrow}(d_{xy})^{\uparrow\downarrow},\\
    E_{g}(b)&:(a_{g})^{\uparrow\downarrow}(e_{g})^{\uparrow\downarrow\uparrow}(b_{2g})^{\uparrow}
    = (d_{z^2})^{\uparrow\downarrow}(d_{xz,yz})^{\uparrow\downarrow\uparrow}(d_{xy})^{\uparrow}.
  \end{split}
  \label{eq:config}
\end{equation}
Any occupations to $d_{x^2-y^2}$ are excluded from the possibility 
because the orbital makes a strong $\sigma^*$-coupling with 
neighboring ligands to get unstabilized.
\section*{Results}
\label{sec:results}
\subsection*{DMC results}
Predictions of the relative stability
among the states are shown in Fig.~\ref{DFT}, 
compared with each other among DMC, CASSCF, and DFT. 
A pronounced feature of the present DMC 
predictions (shown as bold lines) is 
the 'N-shaped' dependence [the lowest $A_{2g}$ and 
a dip at $E_g$(a)]. 
Since CASSCF-DMC is reported to be '{\it not best}'
in the nodal quality sense \cite{2016FUM}, 
we also performed several DFT-DMC results 
using M06, M06L, and M06-2X functionals for 
further confirmation. 
We see that all the DMC give the N-shape with 
the lowest energy by $A_{2g}$, being consistent 
with each other. 
For M06L, it is interesting to see that 
 'non-N-shape' at DFT level 
turns to 'N-shape' at fixed-node DMC level. 
Among the DFT-DMC, the M06 is found to give 
{\it variationaly best} nodal surfaces, making 
total energies of each state lower than 
those by M06L and M06-2X by around $\sim 0.4$ eV in average. 
We cannot make unfortunately the variational comparison 
between DFT-DMC and CASSCF-DMC
because the latter is an all-electron simulation,
while the former a pseudo potential one
(See Supplementary Information for computational details).
%

\subsection*{Inconsistency of DFT predictions }
While most of preceding DFT reported $A_{2g}$ 
predicted as the ground state, 
we clearly see here again that the prediction 
indeed depends on the choice of XC, as already
pointed out by several authors.~\cite{1991REY,2009MAR,2009KUZ} 
Assuming the results by M06-DMC  
being the more reliable prediction, 
we can focus on its 'N-shaped' dependence 
being a target property to be reproduced 
by properly selected XC. 
We see that the B3LYP-DFT prediction is quite similar
to the M06-DMC one, which would support
the validation of DFT predictions by 
this functional to some extent. 
We could identify that the short-range exchange 
contribution is quite essential to reproduce 
the 'N-shape', as discussed below. 

\begin{figure}
  \centering
  \includegraphics[width=\hsize]{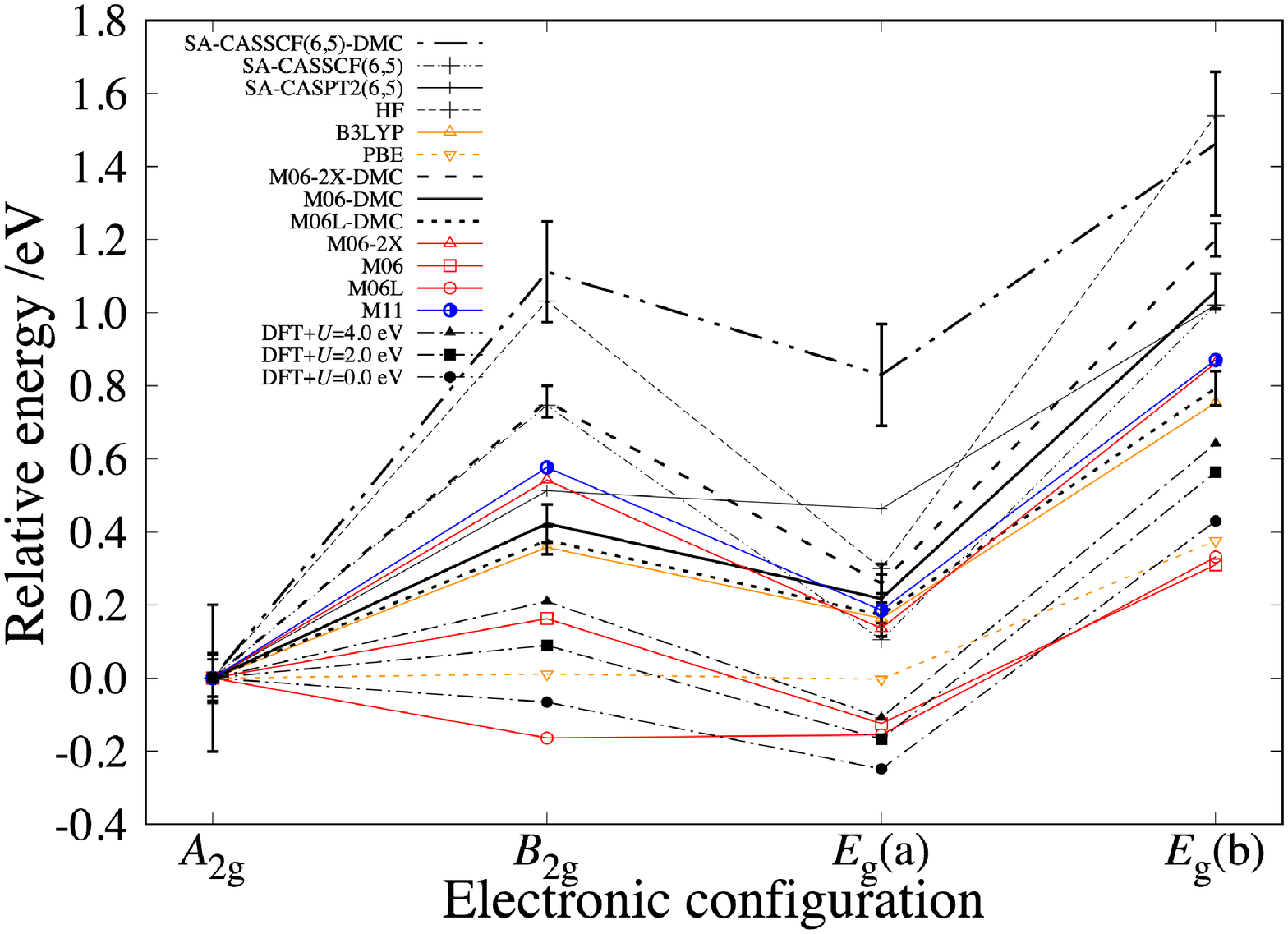}
  \caption{
    The predictions of some {\it ab initio} methods. 
    This graph shows the predictions of CASSCF-DMC, 
    DFT-DMC, 
    HF, CASSCF, CASPT2, and some DFT calculations
    about the relative stability
    among the four electronic configurations.
  }
  \label{DFT}
\end{figure}

Looking at DFT+$U$  results,~\cite{2014HIM}
we see that the 'N-shape' gets recovered as 
$U$ increases, corresponding to taking into 
account the exchange component more. 
The same tendency can also be confirmed 
by the results~\cite{2014PEV} using a hybrid 
functionals, M06. 
In the series of M06 in Fig.~\ref{DFT}, 
the Fock exchange contribution increases as 
M06-L(0\%) $\to$ M06(27\%) $\to$ M06-2X(54\%), 
and it approaches to the 'N-shaped' dependence. 
M11 in Fig.~\ref{DFT} also includes the Fock 
exchange contribution but it is separated into 
short- and long-range components. 
The comparison between M11 and M06-2X clearly shows 
us which component matters: 
The two functionals both contain the Fock exchange 
at almost the same fraction in total, 
54\% (M06-2X) and 42.8\% (M11), 
but they are different in the their fractions of 
long-range component, 
54\% (M06-2X) and 100\% (M11). 
Almost the same 'N-shape' for both implies that
the improvement in M11 from M06-2X
(taking into account the long-range
exchange components) does not so matter
in the present case to reproduce the 'N-shape'.
Though the improvement is known to affect a lot
in reproducing long-ranged natures such as 
van der Waals interactions,~\cite{2014TSU}
what matters in the present case seems rather 
the short-range component, or to say the
self-interaction nature of the exchange.

\section*{Discussions}

\subsection*{Short-range exchange in DFT}
The implication of the importance of
the short-range exchange component
is in accordance with the fact that
the increase of $U$ in DFT+$U$ enhances
the 'N-shape'. 
This would be encouraging to omit
the costly evaluation of long-range exchange 
when one investigates more realistic periodic molecular
crystals.
Since the exchange interaction represents 
the energy gain by the orbital overlap
only for spin-parallel pairs,
it would critically affect when
we estimate the energy differences between
the states with different spin-pair configurations.
While in DMC such factors are taken into account
by default, DFT treatments require the special attention
to choose XC functional so that it could include 
enough short-range exchange component to describe
proper trends in energy differences. 


As mentioned in the introduction,
it is difficult to find such experiments those 
are exactly realizing the isolated/non-relativistic 
molecular system to be compared with the present
theoretical estimations. 
Photoelectron spectroscopy of 
FePc in gas phase \cite{2011BRE} can be the 
nearest case for this purpose.
Another earlier spectroscopy was reported \cite{1974STI}
for the molecule in solvent, concluding $A_{2g}$ as its
ground state configuration because only
this configuration can explain the observed shape of
spectrum. 
A recent spectroscopic study~\cite{2011BRE}
indicates B3LYP agrees reasonably well with experiment
while PBE does not.
This is rather consistent with our trend 
in Fig.~\ref{DFT}, where PBE fails to reproduce
the N-shape while B3LYP with
exchange components can do it.
These consistencies would support our conclusion
that the N-shape by DMC describes
the correct order of the energetic stability
for each of the electronic configurations.

\subsection*{Exchange v.s. Correlation effects}
  Static correlation describing multi-reference nature
  is captured by CASSCF beyond HF,
  while dynamical correlation describing
  hybridization between iron and ligands by CASPT2,
  though the second-order perturbation theory
  is well known to overestimate the dynamic one
for open shell system\cite{2004GHI}.
The corresponding change in Fig.~\ref{DFT} 
can then be identified as a consequence 
of the correlation effect, 
which decreases relative energy differences 
among the configurations making the 
N-shape less pronounced. 
In contrast,
  exchange makes the shape more pronounced.
We could then identify the ratio, 
$\gamma$ = (correlation/exchange), would be a factor 
to dominate the N-shape. 
Ratios for M06L and PBE get larger
owing to incomplete inclusion of exchange part,
and then the shape would is less pronounced.
A comparison between M06 and B3LYP 
would lead to an assumption that 
the shape would be dominated by the ratio rather 
than the absolute intensity of exchange, 
because their 
  inclusion percentages of Fock exchange term are 
almost the same (27\% and 25\%) 
but their shapes considerably differ from each other.
This might be attributed to the 
difference of the {\it ratio} $\gamma$.

\vspace{2mm}
Table~\ref{tab.casscf} lists
  CASSCF expansion coefficients,
  which can be roughly considered
  extent of multi-reference nature.
  According to these values, $A_{2g}$ has
  a less multi-reference nature than the others.
This is just corresponding to the fact 
that the {\it amplitude} of the dependence in 
Fig.~\ref{DFT} gets reduced by CASSCF compared 
with HF. 
The reduction can be explained by the larger 
energy stabilizations for each configuration 
than in $A_{2g}$, those are due to 
the static correlations by 
more enhanced multi-reference nature, 
leading to smaller energy differences 
between each configuration than in HF.

 \begin{table}[htbp]
   \caption{
     The multi determinants calculated by SA-CASSCF
     for each electronic configurations.
     This table shows the occupation of $d$ shell
     of each determinant and its coefficient.
     $\uparrow$ and $\downarrow$ means up spin
     and down spin respectively.
     Only the dominating determinants are listed up,
     whose coefficient's absolute values are higher
     than 0.2.
  }
  \label{tab.casscf}
  \begin{center}
    \begin{tabular}{ccccccc}
      \hline
      state & coefficient  & $3d_{xy}$ & $3d_{xz}$ & $3d_{yz}$ & $d_{z^2}$ & $d_{x^2-y^2}$ \\
      \hline
      $A_{2g}$ &  0.985 & $\uparrow\downarrow$ & $\uparrow$ & $\uparrow$ & $\uparrow\downarrow$ & \\
      $B_{2g}$ &  0.924 & $\uparrow\downarrow$ & $\uparrow\downarrow$ & $\uparrow$ & $\uparrow$ & \\  
              & -0.345 & $\uparrow$ & $\uparrow$ & $\uparrow\downarrow$ & $\uparrow\downarrow$ & \\
      $E_g$(a) &  0.941 &  $\uparrow\downarrow$ & $\uparrow\downarrow$ & $\uparrow$ & $\uparrow$ & \\
               & -0.239 &  $\uparrow\downarrow$ & $\uparrow$  & $\uparrow$ &  $\uparrow$ & $\downarrow$ \\
      $E_g$(a) &  0.778 &  $\uparrow$ & $\uparrow\downarrow$ & $\uparrow$ &  $\uparrow\downarrow$ & \\
               & -0.382 &  $\uparrow$ & $\uparrow\downarrow$ & $\uparrow$ &  $\uparrow$  & $\downarrow$ \\
      \hline
    \end{tabular}
  \end{center}
\end{table}

\vspace{2mm}
Though we cannot make clear statements on
how the ratio $\gamma$ in XC affects 
the qualities of nodal surfaces in FN-DMC, 
but we can see that the {\it amplitude} of the dependence 
in Fig. \ref{DFT} gets larger in the order 
as M06L $\to$ M06 (variationally best among DFT-DMC) 
$\to$ M06-2X getting closer to CASSCF-DMC. 
This trend can be regarded to be driven by 
the decrease of $\gamma$ for the nodal surface 
generation by DFT. 
The {\it location} of CASSCF-DMC at the end of 
this trend would be consistent with the 
preceding report \cite{2016FUM} in the sense that 
the incomplete inclusion of (short-range) correlations 
in CASSCF leads to the diminished $\gamma$ for 
nodal surface generation. 
%
%

\subsection*{Comparison with superposition model}
The present DMC again supports the preceding 
DFT predictions of $^3A_{2g}$ ground state. 
Such a possibility is, however, not permitted 
by the recent 'ligand field'-based 
model,~\cite{2013KUZ} as shown in Fig.~3 
in their paper. 
We can explain this apparent contradiction 
by examining the assumptions made in the models. 
Key is the reduction of number of ligand parameters 
to specify the ground state in the model. 
Original ligand field model for $D_{4h}$ requires 
three parameters, $(D_q, D_s, D_t)$, to specify it, 
as in the studies,~\cite{2009MIE,2015FER}
where $^3A_{2g}$ is still in the possibility 
for the ground state ({\it e.g.} in Fig.~2 in 
the paper by Miedema {\it et al.}\cite{2009MIE}). 
In the recent study by Kuz'min {\it et al.},~\cite{2013KUZ}
however, the ground state is specified only by 
two ligand parameters under the constraint, 
$D_t = (2/35)\cdot 10D_q$, denying the 
$^3A_{2g}$ possibility. 
The constraint comes from the more restricted 
geometry than $D_{4h}$ under the assumptions 
by the superposition model:~\cite{1989NEW}
It assumes that (a) only the nearest neighboring 
ligands (N in this case) are considered, 
(b) ligand fields are specified only via 
bond-length dependence, 
(c) Total ligand field is a superposition of 
each ligand contribution, 
(d) Each contribution is assumed being axially symmetric 
around the intervening bond between the ligand and 
central element (Fe-N in this case). 
The assumption makes the system 
perfect square being able to 
put a Fe-N bond as $x$ (or $y$) axis to 
get more constraint, as explained in 
Kuz'min {\it et al.}.~\cite{2013KUZ}
We thus notice that the $^3A_{2g}$ possibility 
has disappeared by this assumption.

\begin{figure}[htbp]
  \centering
  \includegraphics[width=\hsize]{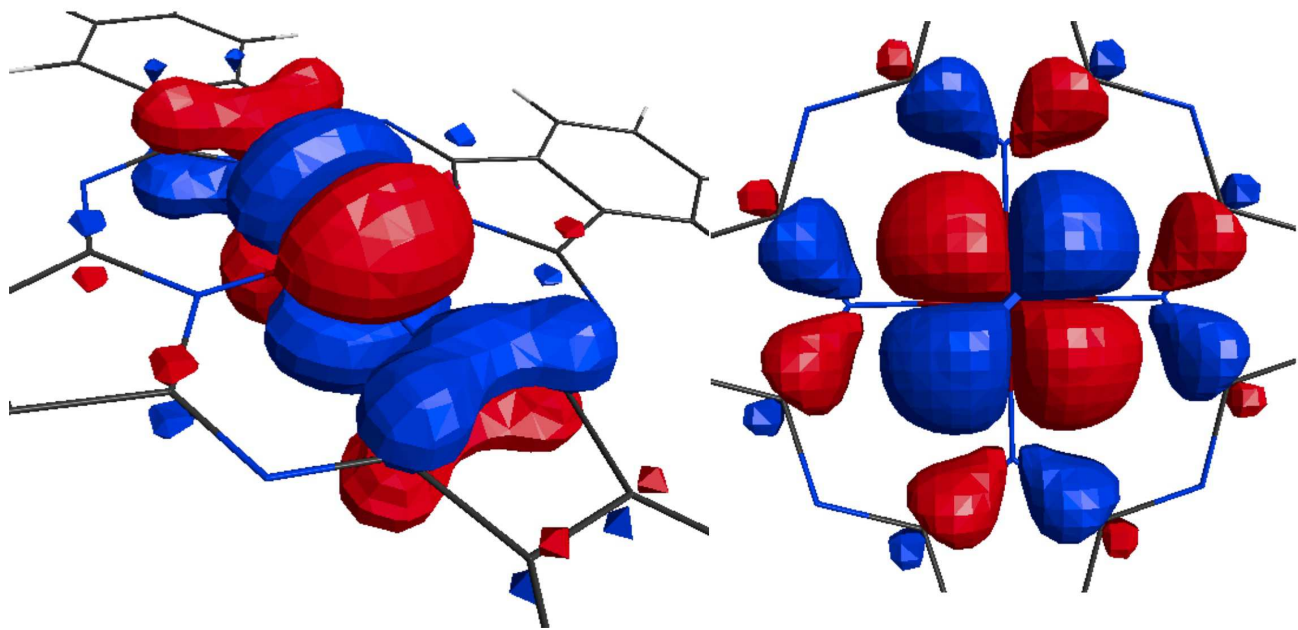}
  \caption{
    The figures of 3d orbitals. $e_{g}$ (left) and $b_{2g}$ (right) orbitals are evaluated by SA-CASSCF(6,5).
  }
  \label{orbitals}
\end{figure}

The assumption, especially (d), seems 
to be easily refuted by the {\it ab initio} 
analysis for realistic treatments of 
actual ligands. 
The left panel in Fig.~\ref{orbitals} shows the 
shape of $e_g$ evaluated by SA-CASSCF(6,5), 
being obviously not the case of the assumption (d). 
We can give further possible explanations 
why the model assumption makes the $^3A_{2g}$ 
unstabilized more than the {\it ab initio} prediction:
From the occupations given in Eq.~\eqref{eq:config},
we notice that $^3A_{2g}$ is stabilized by reducing 
the number of the $e_g$ occupation. 
As seen in the left panel (Fig.~\ref{orbitals}), 
the orbital forms strong $\pi$-coupling with 
neighboring N, spreading along the molecular plane. 
The orbital has the opposite phase to that of Fe, 
forming an unstabilized anti-bonding state. 
The $^3A_{2g}$ stabilizes itself by reducing 
such an energy loss made by the $e_g$ occupation. 
This stabilization mechanism cannot be captured 
by the assumptions of superposition model at all. 
Another mechanism would be captured by the 
right panel of Fig.~\ref{orbitals}, where 
$b_{2g}$ orbital stabilizes itself by 
leaking its distribution toward outer ligands, 
which is not taken into account in the model.
The model then describes the spurious confinement 
for the electrons in $b_{2g}$ orbital, getting 
its energy level increased than the {\it ab initio} 
estimation.
This would also underestimate the stabilization 
of $^3A_{2g}$ via $b_{2g}$ occupation, and hence 
the state has disappeared from the possibility 
of the most stable state.

As a further possible explanation  
for the stabilization of $A_{2g}$ 
as well as that of $E_g$(a), 
we could take the energy loss due to 
the double occupancy on the spatially 
localized orbitals. 
Counting the double occupancies of  
$a_g$/$e_g$/$b_{2g}$ orbitals in 
each state, it is 
1/0/1 [$A_{2g}$], 0/2/0 [$B_{2g}$], 0/1/1 [$E_g$(a)], 
and 1/1/0 [$E_g$(b)], corresponding to a problem 
how to assign two double occupancies to each orbital. 
When the $U$ is introduced, the system wants 
to avoid the double occupancy in a localized orbital,
and then prefers to put it on $b_{2g}$ with 
more spreading $d_{xy}$. 
The states having a $d_{xy}$ double occupancy 
then stabilized relatively to get 
'N-shaped' dependence in Fig.~\ref{DFT}. 
The further stabilization of $A_{2g}$ over $E_g$(a) would be 
explained when we look at a triplet spin pair 
assigned in which orbital.
In $A_{2g}$, the pair is between $d_{zx}$ and $d_{yz}$ 
which are degenerated, while in $E_g$(a) they are between 
$d_{z^2}$ and one of the $d_{zx,yz}$. 
The exchange energy gain is expected to be larger 
when the pair is within the degenerated orbitals, 
making a possible explanation for the stabilization.

\section*{Conclusion}
To conclude, we have applied DMC with the SA-CASSCF
and DFT (M06, M06L and M06-2X) nodes to evaluate relative
stability of the possible electronic configurations
of an isolated FePc under $D_{4h}$ symmetry.
All the DMC simulations predict the ground state to be $A_{2g}$,
supporting several preceding DFT results,~\cite{2001LIA,2009MAR,2012NAK} 
though they have been regarded not well convincing 
because of the ambiguity about XC, as some DFT results predict
different configuration.~\cite{1991REY,2009MAR,2009KUZ}
By making comparisons between 
DMC and several XC, we clarified the importance 
of the short range exchange effect to reproduce 
proper relative stability among the states.
We found M06 gives the variationally best fixed node for DMC.
Interestingly, within the DFT framework,
the B3LYP-DFT prediction was closest to the M06-DMC one.
Getting confidence about the prediction, 
we examined why the predicted $A_{2g}$ is excluded 
from possible ground states in the recent 
ligand field based model.~\cite{2013KUZ}
The assumptions to simplify the model 
are identified to give unreasonably 
less energy gain for $A_{2g}$ 
when compared with the reality. 
The state is found to have possible reasons 
for the stabilization, 
reducing the occupations from an unstable 
anti-bonding orbital, preventing double 
occupancies in a spatially localized 
orbital, and gaining exchange energy 
by putting a triplet spin pair into
degenerated orbitals.

FePc is a typical molecule of MN$_4$ macrocycles
(M = transition metal), which show a promising electrochemical
catalytic activity for the reduction of molecular oxygen. 
The binding of molecular oxygen to the MN$_4$ catalyst involves
binding to the \textit{d}-orbitals of the central metal
in the macrocyclic structure and will be influenced by the
electronic density located on those orbitals.\cite{Zagal2006book}
To the best of our knowledge, this is the first work
which takes into account the many-body wavefunctions
to determine unambiguously the ground state of
3\textit{d}-orbitals in FePc molecule.
We believe that our results would provide useful hints 
about
understanding the interaction of O$_2$ molecules
with active sites in FePc-based catalysts.
As mentioned above, the short range exchange interaction is
very important to describe the relative stability of
different states of FePc molecule.
This provides an important insight into the choice of XC
in {\it ab initio} molecular dynamics studies \cite{1985CAR} on 
oxygen reduction mechanism in FeN$_4$ macrocycles.

\section*{Acknowledgments}
We are grateful to Prof. Kiyo Terakura for leading us
to this challenging topic, to Dr. Yukio Kawashima
for his advice on the generation of CASSCF wavefunctions,
and to Prof. E.D. Jemmis for his enlightening comments.
We also thank the Computational Materials Science Initiative (CMSI/Japan) 
for the computational resources, SR16000 (Center for Computational Materials Science of the 
Institute for Materials Research, Tohoku University/Japan) and K-computer (Riken/Japan).
The computation in this work has been partially performing using
the facilities of the Center for Information Science in JAIST.
K.H. is grateful for financial support from a KAKENHI grant (15K21023), 
a Grant-in-Aid for Scientific Research on Innovative Areas (16H06439), 
PRESTO and the Materials research by Information Integration Initiative (MI$^2$I) project 
of the Support Program for Starting Up Innovation Hub 
from Japan Science and Technology Agency (JST).
R.M. is grateful for financial support from MEXT-KAKENHI 
grants 26287063 and that from the Asahi glass Foundation.
\section*{Author contributions}
\label{sec:author_contribution}
T.I. initiated and performed main calculations under the supervision by
K.H. and R.M., and Z.H. performed DFT+$U$ calculations.
Data is analyzed by all the authors.
All the authors contributed the paper, section by section, 
finally organized to a manuscript.
\section*{Additional information}
Competing financial interests: The authors declare no competing
financial interests.
\newpage
\section*{Supplemetary Information}
\subsection*{SA-CASSCF calculation}
We performed CASSCF calculations using GAMESS
(ver. 5DEC2014R1)\cite{1993SCH, 2005GOR}
to generate the multi-determinant wave functions.
There have been several DMC studies applied to the systems 
including transition metal elements
with the fixed nodes generated by HF or post HF methods.
\cite{2012HON,2008KOS}
Evaluations of energy differences by DMC 
include the study by ROHF-DMC~\cite{2004ASP}, estimating the 
difference between the first and second excited states
of free base porphyrin, getting 
deviations within 0.1~eV from experimental values. 
Dubecky {\it et al.}~\cite{2010DUB} evaluated 
the first excitation energies of 
cis- (trans-) azobenzene using CASSCF-DMC, 
getting the deviation within 0.3 (0.01)~eV. 
Zimmerman {\it et al.}~\cite{2009ZIM} applied it to 
methylene to get deviations within 0.3~eV 
for differences between electronic configurations 
regardless of the active space sizes.

There are two options of CASSCF, 
state-specific CASSCF~(SS-CASSCF) and 
state-averaged CASSCF~(SA-CASSCF). 
In SA-CASSCF, common orbitals are applied to 
all the states to be evaluated while 
SS-CASSCF uses different ones optimized for 
each state individually. 
For the purpose to evaluate energy differences 
used as CASSCF-DMC, SA-CASSCF is known to be 
appropriate choice:~\cite{2009BOU,2012TOU}
For excitation energies of an acrolein molecule, 
SS-CASSCF and SA-CASSCF trial nodes are 
compared in DMC to get the conclusion that 
only SA-CASSCF gives reasonable estimations, 
~\cite{2009BOU,2012TOU}
probably due to the better error cancellations.
Based on that, we generated trial nodes by SA-CASSCF.

All the elements are described as 'all-electron', 
getting the total number of electrons being 290. 
The molecular orbitals are expanded by 6-31G** 
Gaussian basis sets. As a nature of SA-CASSCF, 
we had to use the same geometries commonly for all the states. 
The justifying discussions are given later.
The size of active space is taken as CAS(6,5) 
because it is almost the tractable limit within 
the available computational resources, especially 
for SA-CASSCF. We also performed SA-CASPT2 with the
same conditions.


\subsection*{DFT calculations}


  We performed DFT calculations using Gaussian09~\cite{g09}.
  We run all-electron calculations with def2QZVP basis set
  to compare DFT results. We used Burkatzki pseudo potentials
  with triple-$\zeta$ valence basis set to generate
  the trial nodes for DMC.~\cite{2016BUR}

  To get a symmetry-adapted state for each electronic configuration
  in a DFT simulation, we used "guess=alter" and
  "scf=symm" implemented in Gaussian09:
  We first give an initial guess appropriate for the target state
  by "guess=alter", and then fix the symmetries of the all occupied orbitals
  during its SCF procedure by "scf=symm".

\subsection*{DFT+U calculations}
We have performed the DFT+U calculations for FePc molecule
by using a simplified version of Cococcioni and
de Gironcoli~\cite{Cococcioni2005prb}, as implemented
in QUANTUM ESPRESSO package~\cite{Giannozzi2009jpcm}.
Several different values (0, 2 and 4 eV) have been considered
for Hubbard $U$ parameter for Fe 3$d$ orbitals.
We have employed ultrasoft pseudopotentials generated
with the Rappe-Rabe-Kaxiras-Joannopoulos recipe~\cite{Rappe1990prb}
to represent electron-ion interaction.
The electronic exchange-correlation potential was calculated
within the generalized gradient approximation (GGA)
using the scheme of Perdew-Burke-Ernzerhof (PBE)~\cite{Perdew1996prl}
and the spin-polarization was taken into account.
The electronic wave functions were expanded in plane waves
with an energy cutoff of 35 Ry while for the charge density the energy cutoff
was taken to 350 Ry. The isolated FePc molecule was simulated
in a simple tetragonal cell of $27 \times 27 \times 12$ \AA$^3$.
Brillouin-zone integrations were approximated using a $\Gamma$ point.
The atomic positions of FePc molecule were optimized till the residual
forces were less than 0.01 eV/\AA.


\subsection*{Jastrow factor}
We adopted a Jastrow factor~\cite{2004DRU} 
multiplied by determinant(s)
to form a guiding function for DMC, 
imposing Kato's cusp conditions.~\cite{1957KAT}
We used a function form implemented in CASINO~\cite{2010NEE} 
including electron-electron ($u$), 
electron-nuclei ($\chi$), and electron-electron-nuclei ($f$) 
terms.
Considering spin polarizations, $u$ and $\chi$ ($f$) 
terms are expanded upto 8th (2nd) order of the power 
of inter-particle distances, getting total 144 
variational parameters. 
Cutoff lengths for these terms are fixed as the 
recommended values by the implementation, and 
all the linear variational parameters are 
optimized by 'varmin-linjas' scheme.~\cite{2005DRU}
For all-electron DMC, we also used the cusp-correction 
scheme~\cite{2005MA} for possible electron-nuclei 
coalescence.
  All the present QMC simulations were done using
  CASINO (ver.~2.13)~\cite{2010NEE}.

\subsection*{Extrapolation of time step error}

For time-step error corrections in DMC,~\cite{2001FOU}
we used an extrapolation scheme~\cite{2011LEE}
using two time steps, $\tau_{1}=\tau_{max}$ and 
$\tau_{2}=\tau_{max}/4$. 
$\tau_{max}$ is taken so that it is 
the largest possible below which the error is 
proportional to $\tau$.
From the results, $E_j \pm \sigma_j$ 
by $\tau_{j}$ ($j=1,2$), the extrapolation is 
evaluated as, 
$E(\tau \to 0) = \left({{{\tau_2}{E_1} - {\tau_1}{E_2}}}\right)/\left({{{\tau _2} - {\tau _1}}}\right)$
and
  $E(\tau \to 0)
  = \left({{{\tau_2}{E_1} - {\tau_1}{E_2}}}\right)/\left({{{\tau _2} - {\tau _1}}}\right)$ and 
  ${\sigma \tau \to 0)} = \left({{\sigma_1^2\tau _2^2
      + \sigma_2^2\tau _1^2}}\right)/{{{{\left( {{\tau _2} - {\tau _1}} \right)}^2}}}$.

For all electron calculations,
It is proposed~\cite{2011LEE} to take 
$\tau_{max}<1/(3Z^2 )$ where $Z$ is the 
maximum atomic number within the system. 
Using $Z=26$ for the present case, we 
chose ${\tau _{\max }} = 4.0 \times {10^{-4}} a.u.$
for SA-CASSCF-DMC. 
The choice actually proved to work except $E_g(a)$ 
as discussed later.
The best practice is known~\cite{2011LEE} 
to accumulate eight times larger steps for
$\tau_{2}$ than that for $\tau_{1}$ to
minimize computational costs. 

For pseudo potential calculations,
$\tau_1=4.0\times{10^{-3}}$ and $\tau_2=1.0\times{10^{-3}}$
were chosen for DFT-DMC.
These values are larger than those of SA-CASCF-DMC.
A necessary resolution in time step is larger in the pseudo potential
case than in the all-electron because random walkers diffuse on
shallower potentials.
  
\begin{figure}
  \centering
  \includegraphics[width=\hsize]{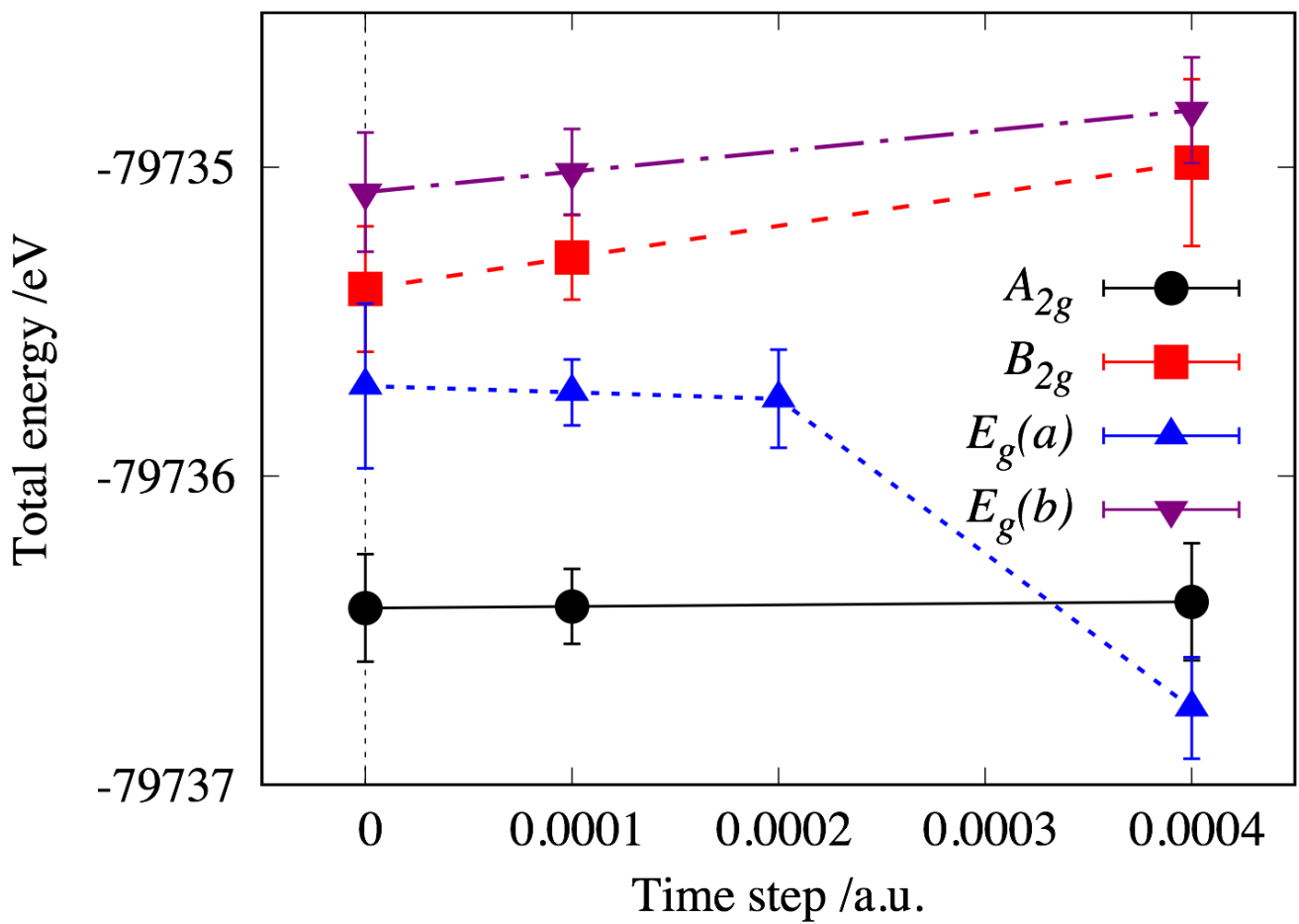}
  \caption{
    The extrapolation of DMC results.
    We can identify the ground state as $A_{2g}$ 
    with 1$\sigma$ statistical confidence, while 
    we cannot for other excited states. 
    Except $E_{g}(a)$, linear extrapolations below 
    $\tau_{\rm max}=0.0004$ works well.
    For $E_{g}(a)$, we extrapolate using 
    $\tau_{\rm max}=0.0002$, instead.
  }
  \label{DMC}
\end{figure}

  \begin{table}[htbp]
  \caption{
    The energy gains and the bond length owing to geometry optimization.
    B3LYP-DFT estimations of geometry relaxation gains, $\Delta E$, 
    and changes in the Fe-N bond lengths ($R_\mathrm{Fe-N}$). 
    $\Delta E$ is defined as the gain when the 
    geometries are optimized from the common structure 
    ~\cite{2008SUM} used in the present SA-CASSCF. 
    For each $E_g$ state, two bond lengths are given 
    because it falls into $D_{2h}$ from $D_{4h}$ 
    by the relaxation.
  }
  \label{B3LYP}
  \begin{center}
    \begin{tabular}{ccccc}
      \hline
      \hline
      state & & $\Delta E_g$ (meV) & & $R_\mathrm{Fe-N}$ (\AA) \\ 
      \hline
      $A_{2g}$   & & $-0.0$           & & $1.949$ \\ 
      $B_{2g}$   & & $-4.7$ ($-1.3$~\%) & & $1.946$ \\ 
      $E_{g}(a)$ & & $-1.1$ ($-1.2$~\%) & & $1.946/1.950$ \\ 
      $E_{g}(b)$ & & $-7.7$ ($-1.0$~\%) & & $1.941/1.951$ \\ 
      \hline
      \hline
    \end{tabular}
  \end{center}
\end{table}

\subsection*{Effects of geometry differences}
As mentioned above,
SA-CASSCF-DMC restricts us to use the same 
geometry~\cite{2008SUM}
to all the states of electronic configurations. 
When the optimized geometry for each state 
largely differs from each other, this restriction 
could make the estimation poor. 
SA-CASSCF-DMC for an acrolein molecule~\cite{2009BOU}
seems to be the case that it gave the overestimation 
of the excitation energy by $\sim$150~meV: 
Though it is not explicitly stated in their 
paper,~\cite{2009BOU} 
the bond length between carbon and oxygen 
gets elongated by $8~\%$\cite{1960INU,1983BLO} 
when the system is excited. 
Because of the restriction, however,  
SA-CASSCF-DMC cannot take into account the 
relaxation energy gain by the elongation, 
and this is quite likely to be an origin of 
the overestimation. 
To examine if this matters in the present case, 
we evaluated the energy gains by the relaxation 
from the geometry used in SA-CASSCF, 
as tabulated in Tab.~\ref{B3LYP}.
It is confirmed that the gains remain within 
1.3~\%, corresponding to 7.7~meV which is 
negligibly small compared to the statistical 
errors and to the energy scale in Fig.~1 in
the paper. 
The largest relaxation is found to occur on the 
bond between iron and neighboring nitrogen, 
which is confirmed to be within $0.26~\%$ at most. 
The geometry insensitivity to the occupations 
to be considered is, incidentally, in accordance 
with a report~\cite{1998CHO} that the 
Fe-N bonding length is mainly dominated by the 
occupation number of $d_{x^2-y^2}$, which is 
not considered here, though the conclusion 
is drawn from Fe porphyrin case. 
The insensitivity could justify the use of 
the same geometry to evaluate the relative 
stabilities among the states.

\end{document}